\documentclass[preprint,aip,jap,superscriptaddress,showpacs,citeautoscript]{revtex4-1}

\usepackage{graphicx}
\usepackage{dcolumn}
\usepackage{bm}
\usepackage{flafter}
\usepackage{lettrine}
\usepackage{amsmath}
\usepackage{amssymb}
\usepackage{color}
\usepackage{epsfig}

\usepackage{type1cm}
\usepackage{tabu}

\graphicspath{{eps+pdf/}}

\begin{document}

\title{Synthesis of Xenon and Iron/Nickel intermetallic compounds at Earth's core thermodynamic conditions}

\author{Elissaios Stavrou }
\email[E-mail E.S. ]{stavrou1@llnl.gov}
\affiliation {Lawrence Livermore National Laboratory, Physical and Life Sciences Directorate,
Livermore, California 94550, USA}
\author{Yansun Yao}
\email[E-mail Y.Y. ]{yansun.yao@usask.ca}
\affiliation {Department of Physics and Engineering Physics, University of Saskatchewan, Saskatoon Saskatchewan, S7N 5E2, Canada}
\affiliation {Canadian Light Source, Saskatoon, Saskatchewan, S7N 2V3, Canada }
\author {Alexander F. Goncharov}
\email[E-mail A.G. ]{agoncharov@carnegiescience.edu}
\affiliation{Key Laboratory of Materials Physics and Center for Energy Matter in Extreme Environments,  Chinese Academy  of Sciences, Hefei 230031, China}
\affiliation{Geophysical Laboratory, Carnegie Institution of Washington, Washington, D.C. 20015, USA}
\affiliation{University of Science and Technology of China, Hefei, 230026, China}
\author{Sergey Lobanov}
\affiliation{Geophysical Laboratory, Carnegie Institution of Washington, Washington, D.C. 20015, USA}
\affiliation{V.S. Sobolev Institute of Geology and Mineralogy, SB RAS,3 Pr. Ac. Koptyga, Novosibirsk 630090, Russia.}
\author{Joseph M. Zaug}
\affiliation {Lawrence Livermore National Laboratory, Physical and Life Sciences Directorate,
Livermore, California 94550, USA}
\author{Hanyu Liu}
\affiliation{Geophysical Laboratory, Carnegie Institution of Washington, Washington, D.C. 20015, USA}
\author{Eran Greenberg}
\affiliation{Center for Advanced Radiation Sources, University of Chicago, Chicago, IL 60637, USA}
\author{Vitali B. Prakapenka}
\affiliation{Center for Advanced Radiation Sources, University of Chicago, Chicago, IL 60637, USA}

\date{\today}

\begin{abstract}
Although Xe is known to form stable compounds with strong electronegative elements,  evidence on the formation of stable compounds with electropositive elements, such as Fe and Ni, was missing until very recently. In addition to the significance of the emerging field of noble gas elements chemistry, the possible formation of Xe-Fe/Ni compounds has been proposed as a plausible explanation of the so-called \textquotedblleft missing Xe paradox\textquotedblright. Here we explore the possible formation of stable compounds in the Xe-Fe/Ni  systems at thermodynamic conditions representative of Earth's core.  Using in situ synchrotron X-ray diffraction and Raman spectroscopy in concert with first principles calculations we demonstrate the synthesis of stable Xe(Fe,Fe/Ni)$_3$ and XeNi$_3$  compounds.  The results indicate the changing chemical properties of elements under extreme conditions where  noble gas elements can form stable compounds with elements which are electropositive at ambient conditions  but become slightly electronegative at high pressures.
\end{abstract}
\maketitle

Noble gas   elements (NGEs) are considered as the most chemically inert elements due to the closed subshells configuration that prevents the formation of stable compounds.  However, recent theoretical studies \cite{Zhu2014,Li2015,Miao2015,Zhang2015} suggest that stable compounds between NGEs and metals (electropositive at ambient conditions elements) could be formed  under high pressure conditions due to the substantial effect of pressure on the chemical properties. The stability of such compounds can be attributed to the changes of chemical properties of elements under pressure \cite{Grochala2007,Zhu2013,Zhu2014,Dong2015}, which includes altered electronegativity and reactivity, charge transfer between orbitals and/or between constitution elements, and  the appearance of  multicenter bonding and electride states as in the case of the recently synthesized Na$_2$He \cite{Dong2017} compound. In general,  for the predicted stable compounds of NGEs and metals a NGE can either  gain electrons from an electropositive, at the corresponding pressures, element (e.g. the Xe-Mg system \cite{Miao2015}) or a metal becomes electronegative, at the corresponding pressures, and acts as an oxidant (e.g. Xe-Fe/Ni system \cite{Zhu2014}). The latter case is unusual and counters chemical intuition because it implies that  Fe and Ni  become more electronegative than Xe. Experimental realization of such compounds is incomplete  highlighting the necessity of experimental verification of theory to better understanding of the chemistry at extreme conditions that is important for advancing chemistry and physics of highly compressed material states.

The formation of stable Xe-Fe(Ni) compounds at pressures comparable to those of the core accretion would also change our understanding about the presence of Xe in the Earth's core. According to the simple mass fractionation model (see discussion in Ref. \cite{Dauphas2003} and references therein), heavy NGEs should be less depleted and isotopically fractionated in comparison to the lighter ones, in agreement to observations in meteorites.  However, in the Earth's atmosphere, Xe is more depleted than Kr and more fractionated than both Kr and Ar \cite{Dauphas2003}. These two observations constitute one of the most challenging open questions in the geosciences \cite{Anders1977,Dauphas2003}, and give rise to the so-called  \textquotedblleft missing Xe paradox\textquotedblright. Although various models have been suggested on the origin of the depletion aspect of the \textquotedblleft missing Xe paradox\textquotedblright \cite{Shcheka2012}, it is commonly attributed to the inclusion of Xe in the Earth's interior\cite{Lee2006}. In this scenario, other NGEs, such as Ar, Kr and Ne, are not \textquotedblleft missing\textquotedblright {} due to the much more extreme thermodynamic conditions needed for the formation of stable compounds with metals \cite{Miao2015}. While Xe was reported to form compounds with water ice \cite{Sanloup2013} and silica \cite{Sanloup2005} at low pressures, none of them provide a plausible explanation to the \textquotedblleft missing Xe paradox\textquotedblright. Indeed, only a very low amount of Xe can be incorporated \cite{Sanloup2005} in silica at $<$ 1 GPa and 500 K while first-principles studies \cite{Probert2010} indicate that substitution of Si by Xe is very energetically unfavorable. The successful formation of xenon oxides under deep mantle conditions has been recently reported \cite{Dewaele2016}. However, the presence of such compounds is precluded by the lack of free oxygen in Earth's mantle. Accordingly, a hypothesis of stable Xe-Fe/Ni compounds  in the Earth's core was proposed as an explanation for the \textquotedblleft missing\textquotedblright {}  Xe \cite{Lee2006}, but no experimental support has been offered.

Previous experimental attempts have failed in  tracing the possible formation of Fe-Xe compounds up to 200 GPa and below 2500 K \cite{Nishio2010, Dewaele2017} and this has been attributed to the large size difference between  Xe and Fe ions, which hinders the formation of substitutional Xe-Fe solid solutions (SS)  according to the Hume-Rothery rule \cite{Nishio2010}. A  recent theoretical study \cite{Zhu2014}, using \emph{ab-initio} calculations combined with multiple structural search methods, suggests that Xe-Ni and Xe-Fe compounds are thermodynamically stable above 150 GPa and 200 GPa respectively. This study also suggests that the thermodynamic stability of these compounds is further enhanced at elevated temperatures ($>$2000 K) \emph{i.e.} at thermodynamic conditions representative of Earth's outer core. The predicted crystal structures of  Xe-Fe/Ni compounds  are distinct from the structures of elemental Xe, Fe, and Ni at the same thermodynamic conditions, which suggests that the formation mechanism of these compounds goes beyond a simple element substitution in parent compounds.

In this study,  we explored the possible formation of stable compounds in the Xe-Fe/Ni system at thermodynamic conditions representative of the Earth's core by performing high pressure experiments in a laser-heated (LH) diamond-anvil cell (DAC) starting from the following mixtures: a)Xe-Fe, b) Xe-Fe/Ni alloy representative of Earth's core ($\sim$ 7\% Ni) and c) Xe-Ni, in an approximate 1:1 loading ratio. Using \emph{in situ} synchrotron X-ray diffraction (XRD) and Raman spectroscopy we successfully identified the formation of: a) a XeFe$_3$/Xe(Fe$_{0.93}$Ni$_{0.07}$)$_3$ compound,  characterized as a mixture of a FCC and an  orthorhombic NbPd$_3$-type structures, above 200 GPa and 2000 K and b) a XeNi$_3$ compound, in the form of a CrNi$_3$-type FCC structure, above 150 GPa and 1500 K. Preliminary data on all these observations have been reported at the AGU 2015 Fall meeting \cite{Stavrou2015}. We find the formation of XeFe$_3$  compounds above 200 GPa (in contrast with previous studies  \cite{Dewaele2017}) while  XeNi$_3$ forms at much lower pressure signifying the importance of the elemental electronic structure on the new compound formation.    The experimental results were examined  and supported in synergy with  theoretical \emph{ab-initio}  structural search and optimization. The formation of XeFe$_3$  and XeNi$_3$ compounds are kinetically driven with the structures identified in  close proximity of the computed energy minima. The theoretical reaction threshold pressures and XRD patterns for both compounds are in very good agreement with the experiment.

\section*{Results and discussion}
In order to study the Xe-Fe/Ni system we   examined  mixtures of Xe with either pure Fe and Ni or with a Fe-Ni alloy with a Ni concentration (7-8\%) representative of the Earth's core \cite{Dubrovinsky2007,Bottke2006}.  In the latter case an iron Sikhote-Alin Meteorite was used as a proxy after chemical and homogeneity characterization using EDX spectroscopy (Fig. S1). The XRD patterns of the  Fe-Ni alloy used in this study are representative of a HCP structure (see Fig 1(a)) with negligible cell volume difference (Fig. S2), at a given pressure, to  that of pure Fe (also in HCP structure) in agreement with previous studies \cite{Mao1990}. We  performed LH experiments on both the Xe-Fe and Xe-Fe$_{0.93}$Ni$_{0.07}$ systems at various pressures from 150 to above 210 GPa. No new Bragg peaks, signalling the formation of  new compounds, were observed below 195 GPa for both mixtures even after a prolonged LH above 3500 K, see Fig.  S3. However,  new Bragg peaks appeared for both mixtures after LH at $>$2000 K and pressures $>$200 GPa, implying an approximately 200 GPa reaction threshold, see Fig. 1(a) and Fig. S4(b). XRD patterns of the Xe-Fe and Xe-Fe$_{0.93}$Ni$_{0.07}$ systems after LH (Fig. S4(a)) are essentially   identical with a slight variations in relative  intensities of Bragg peaks. Thus, we suggest that the presence of a low-concentration  of Ni in the Fe-Ni alloy has no effect on the structure of the synthesized compound. Bragg peaks   of  pure Ni or  a Fe$_{0.97}$Ni$_{0.07}$ BCC structure  \cite{Dubrovinsky2007} were not observed  during or after LH. Thus,  the  possibility of phase separation or a phase transition of the Fe$_{0.97}$Ni$_{0.07}$ alloy are ruled out.

\begin{figure}[ht]
\centering
\includegraphics[width=130mm]{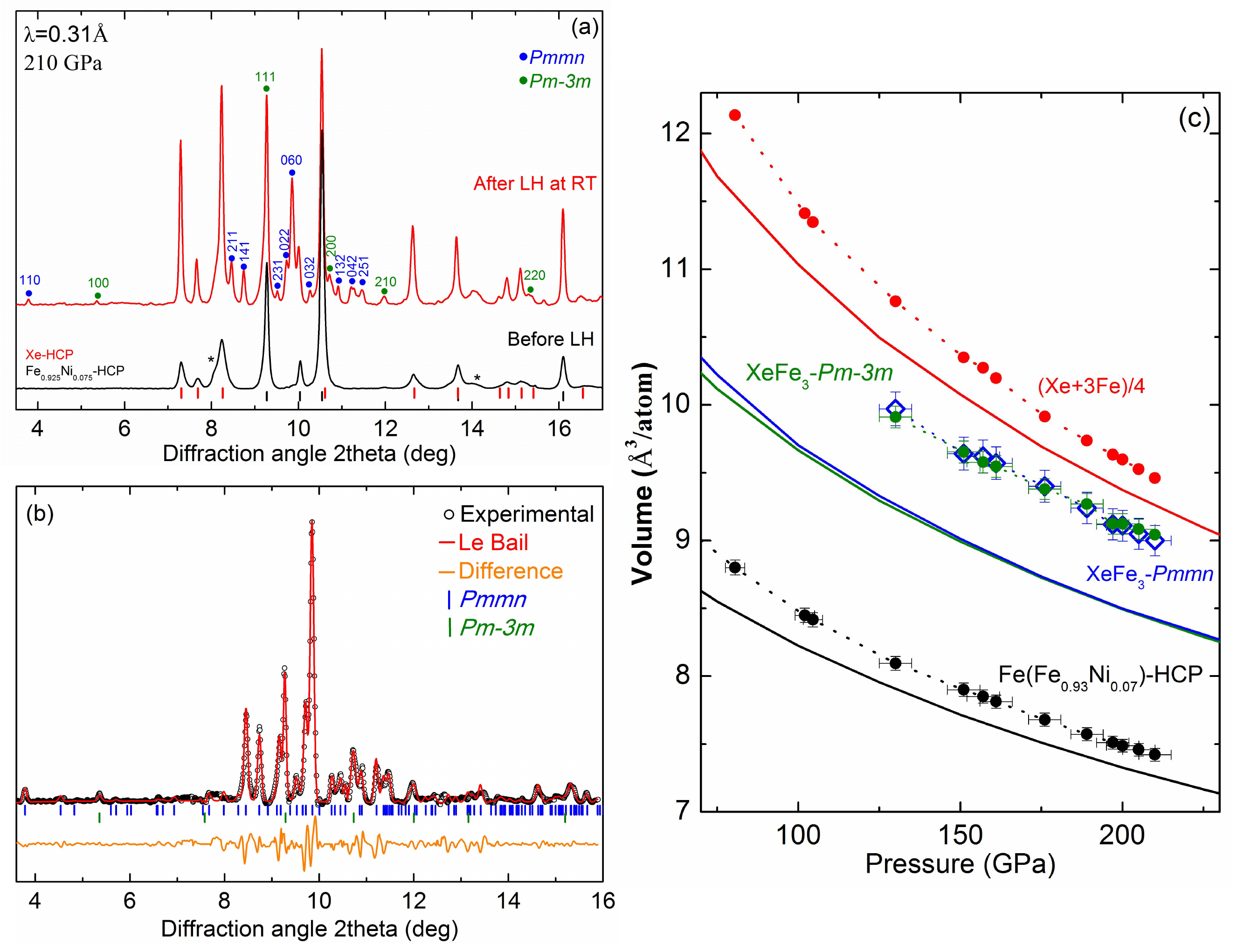}
\caption{X-ray diffraction patterns and equation of state of  XeFe$_3$/Xe(Fe$_{0.93}$Ni$_{0.07}$)$_3$. a) XRD patterns of the  Xe-Fe$_{0.93}$Ni$_{0.07}$ mixture before and after LH at 210 GPa. New Bragg peaks after LH and corresponding Miller  indices for the \emph{Pm-3m}  and the \emph{Pmmn}  XeFe$_3$  crystal structures are noted with green and blue circles respectively.  The peak marked by the asterisk corresponds to the strongest peak of rhenium (gasket material). (b) Le Bail refinement of the Xe(Fe$_{0.93}$Ni$_{0.07}$)$_3$ compound  at 210 GPa.  The peaks of \emph{Pm-3m} and \emph{Pmmn} (1) structures   are marked with green and blue vertical lines, respectively. The difference between the data and the fit is shown below (orange line).  (c) EOSs of  Fe and XeFe$_3$/Xe(Fe$_{0.93}$Ni$_{0.07}$)$_3$  as determined experimentally (dashed curves and solid symbols) and theoretically (solid curves) in this study. The volume of the superposition of (Xe+3Fe)/4 is also shown for comparison.}
\end{figure}

The new peaks in XRD patterns after LH of the Xe-Fe and Xe-Fe$_{0.97}$Ni$_{0.07}$ mixtures cannot be indexed solely with the FCC (\emph{Pm-3m}) XeFe$_3$ structure (Cu$_3$Au type)  by Zhu et al. \cite{Zhu2014} due to a much higher number of observed Bragg peaks and the presence of low angle peaks (see Fig. 1(a)). Moreover, no Raman active vibrational modes are expected for the Cu$_3$Au-type structure in contrast with our Raman spectroscopy measurements (Fig. S5(a)). We identified the products as a mixture of a FCC and an orthorhombic (namely \emph{Pmmn} (1)) phases  with competitive enthalpies as revealed in our theoretical calculations. Details on the  procedure we followed for the identification of the \emph{Pmmn} (1) phase can be found in the Supporting Information. The \emph{Pmmn }(1) and the FCC structures are closely related as both are  close-packed and with 12-fold coordinated Fe and Xe atoms. As a result, the volumes of these two structures are very similar, essentially degenerate at pressures above 100 GPa (Fig. 1(c)).  The Bragg peaks of the experimental XRD patterns can be very well indexed with a mixture of  \emph{Pmmn (1)} and \emph{Pm-3m} structures. However, preferred orientation  effects and strongly anisotropic peak broadening effects, which are usual in HP-HT synthesis \cite{Stavrou2016}, prevent us from a full structural refinement (Rietveld) of the positional parameters due to differences between the observed and calculated intensities. Difference in relative intensities  could also arise from a positional disordered phase. For this reason, we have  considered a positional disordered \emph{Pmmn (1)} structure with a Xe(25\%)-Fe (75\%) site occupancy. This structure has a negligible enthalpy difference with the ordered \emph{Pmmn (1)} one and  provides a better agreement with the experimental XRD patterns. In Fig. 1(b) we show the Le Bail refinement  of the experimentally observed diffraction pattern based on a mixture of \emph{Pmmn} (1) (with Xe(25\%)-Fe (75\%)) and  \emph{Pm-3m} structures, after subtracting(see Ref. \cite{Struzhkin2016}  for the details).

Raman experiments on samples quenched to 300 K (Fig. S5(a)) show the presence of a new broad weak peak at 450- 480 cm$^{-1}$. Low intensity Raman spectra are consistent with the formation of a metallic or semi-metallic XeFe$_3$ compound \cite{Zhu2014} and consequently only the highest intensity peaks are expected to be observed.  The position of the observed Raman peak is indeed in agreement with the strongest calculated peak   of the \emph{Pmmn (1)} XeFe$_3$ phase. Please note that the FCC XeFe$_3$ compound is not expected to have any Raman activity due to the Raman selection rules. Thus the presence of the Raman bands strongly supports the existence of a second \emph{Pmmn} (1)  phase in addition to FCC XeFe$_3$. Moreover,  the pressure slope of the experimentally observed peak and  of the most intense  peak of the  calculated Raman spectrum ( Fig. S5(b))  agree well thus, providing an additional argument in favor to the synthesis of the \emph{Pmmn} (1) XeFe$_3$ phase.

Using the lattice parameters obtained from the XRD patterns, the volume of XeFe$_3$ of the NbPd$_3$-type structure, is ~5\% lower than that of the 1:3 solid mixture of Xe and Fe, suggesting XeFe$_3$ is a stable compound (see Fig. 1(c)). The theoretical EOS yields the same trend, with the XeFe$_3$ having ~ 8\% smaller volume than the mixture. Compared to the experimental values, the theoretical volumes of Fe and XeFe$_3$/Xe(Fe$_{0.93}$Ni$_{0.07}$)$_3$ are clearly underestimated. Here a possible source of error is the well-known insufficiency of standard DFT treating the ground state of Fe, which was shown to be largely affected by dynamical many-body effects \cite{Pourovskii2014}. Nevertheless, experiment and theory agree in that  XeFe$_3$ has a smaller volume than its constituents, suggesting this compound is thermodynamically favored. On pressure release, experiments show that both the XeFe$_3$ and the Xe(Fe$_{0.93}$Ni$_{0.07}$)$_3$ compounds remain stable down to, at least, 127 GPa (Fig. S6).

\begin{figure}[ht]
\centering
\includegraphics[width=130mm]{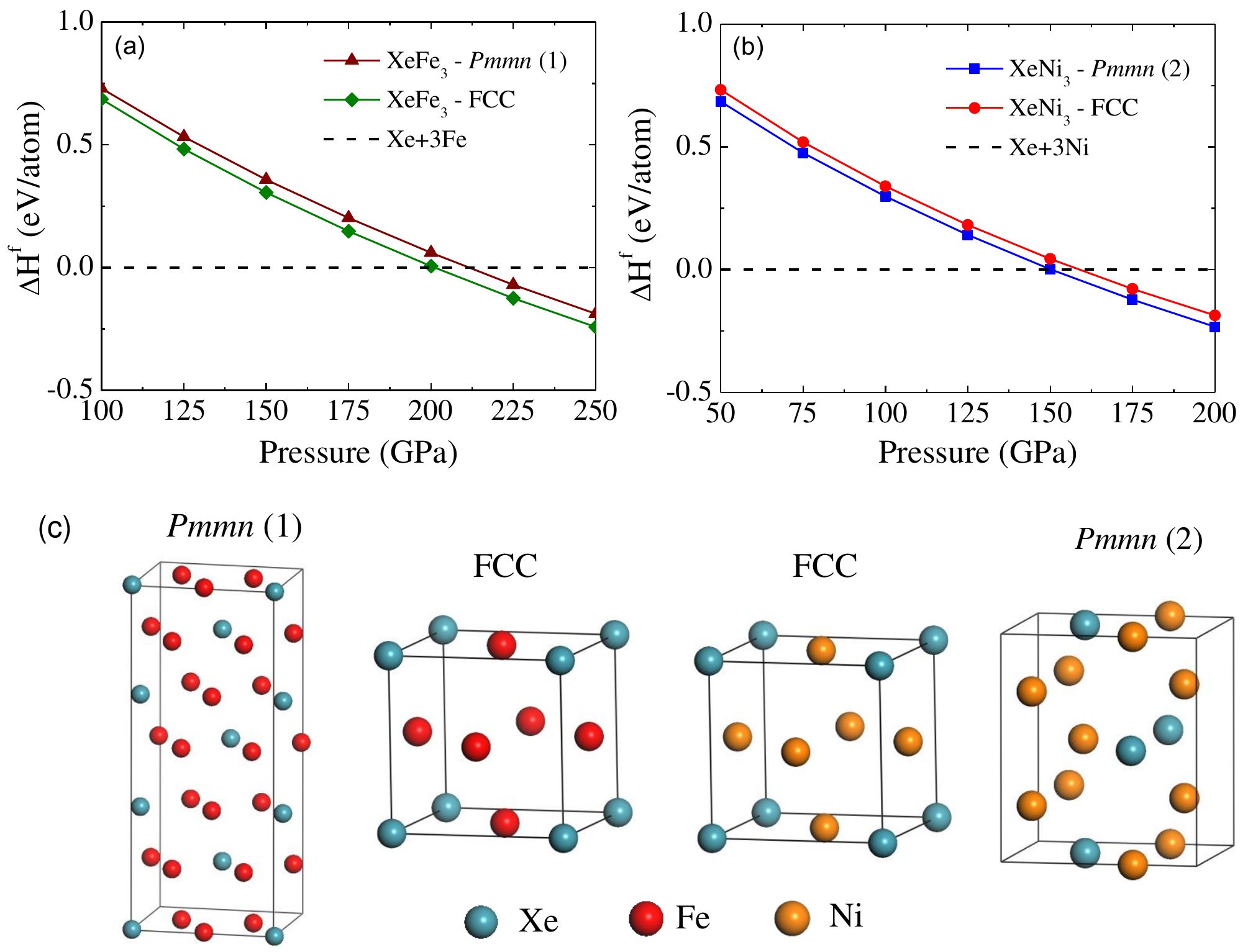}
\caption{Calculated enthalpies of formation  $\Delta$$H^f$  of:  (a)  XeFe$_3$ and (b) XeNi$_3$  with respect to the mixture of elemental Xe + 3Ni and Xe + 3Fe, respectively. The enthalpies of the FCC structure in both compounds were calculated in an ordered structure. The two \emph{Pmmn} structures in XeFe$_3$ and XeNi$_3$  are distinctly different and therefore distinguished as \emph{Pmmn} (1) and \emph{Pmmn} (2). (c) Schematic representations of the corresponding structures of XeFe$_3$ and XeNi$_3$. }
\end{figure}

Figure 3(a) shows  XRD patterns of the Xe-Ni mixture at 155 GPa  before LH, upon increasing temperature   and at RT after LH. The XRD pattern before LH  is representative of a heterogeneous mixture of HCP-Xe \cite{Jephcoat1987} and FCC-Ni.   With increasing temperature  the Ni Bragg peaks completely disappear  above 1500 K  while new peak appear concomitantly suggesting that Ni fully reacts towards the formation of a new compound that remains stable after quenching to RT;  Xe related Bragg peaks remain present  suggesting the conditions of Xe excess in the cavity. The new Bragg peaks can be indexed with an A1 FCC unit-cell and with a cell volume representative of a XeNi$_3$ compound. This attribution is based on   the comparison between the atomic volumes of the synthesized compound, Ni and Xe  at the same pressure (see Fig. S7 and Fig. S8(b)). Fulfillment of the extinction conditions  of A1  by  the observed reflections  implies the formation of a Xe-Ni CrNi$_3$-type  binary alloy with Xe and Ni distributed randomly/statistically over the FCC sites.  An ordered FCC structure (Cu$_3$Au-type) would have several additional low intensity Bragg peaks  (see Fig. 3(b)), which are absent in the XRD pattern of XeNi$_3$.

The synthesized XeNi$_3$ compound remains stable up to at least 100 GPa upon pressure release, see Fig. S8(a). Significantly, both the experimentally determined and calculated volume of XeNi$_3$ is ~10\% smaller than the 1:3 solid mixture of Xe and Ni, suggesting that the former is a stable compound (Fig. S7(b)). However, exact stoichiometry of the synthesized compound may not be precisely determined. Nevertheless,  both the experimentally determined EOS and the predicted stability of the XeNi$_3$ compound strongly suggest a composition very close, if not exact, to XeNi$_3$. The thermodynamic stability of the XeNi$_3$ compound was investigated through the relative enthalpy of formation, $\Delta$$H^f$, with respect to a 1:3 solid mixture of Xe and Ni (Fig. 2(b)). The FCC structure is comparable in enthalpy with the \emph{Pmmn} structure (named here as \emph{Pmmn} (2)) predicted by Zhu et. al. \cite{Zhu2014}. The $\Delta$$H^f$ of the FCC structure is slightly higher than the latter one, \emph{i.e.}, by $\sim$0.04 eV/atom, indicating a metastable structure close to the global minimum. Considering that the formation of XeNi$_3$ only takes place at high temperature it is reasonable to suggest that the synthesis of this compound is kinetically-driven \cite{Ozolins1998, Lu1991}. Remarkably, the $\Delta$$H^f$ of the FCC structure approaches zero near 158 GPa, which corresponds very well with the experimental reaction threshold of 155 GPa.  Both FCC and \emph{Pmmn} (2) structures are close-packed with 12-fold coordinated Ni and Xe atoms (Fig. 2(c)), which explains their similar enthalpies.The enthalpy change due to the positional disorders of Xe and Ni was estimated using a FCC supercell of 256 atoms. A set of 200 structures were generated by placing Xe and Ni atoms randomly at the FCC lattice sites; each representing a possible solid solution configuration.  The calculated enthalpies of these structures (at 150 GPa) is within a 0.1 eV/atom range above the enthalpy of the ordered FCC structure.

Recently, Dewaele \emph{et al.} \cite{Dewaele2017} reported the synthesis of a stable XeNi$_3$ compound with an ordered FCC  (Cu$_3$Au-type) structure. Although the reported stoichiometry,  the reaction threshold, the volume per atom   and the fundamental crystal structure (FCC-type) are in agrement with this work (see also Ref. \cite{Stavrou2015})   a  discrepancy exists on the detailed crystal structure \emph{i.e.} ordered vs disordered FCC. We speculate that this can be attributed to  differences in the  quenching time. The formation of an ordered structure requires substantial atomic diffusion, which is likely restricted by the fast kinetics in the present case, \emph{i.e.}, the quenching process. Strictly speaking, in a positional disordered structure the volume is a statistical average, which may deviate slightly from that of an ordered structure. The present calculation reveals that the deviation is negligible in the present thermodynamic scale which is further justified by the agreement of the reported experimental volumes per atom (Fig. S7 ).

\begin{figure}[ht]
\centering
\includegraphics[width=130mm]{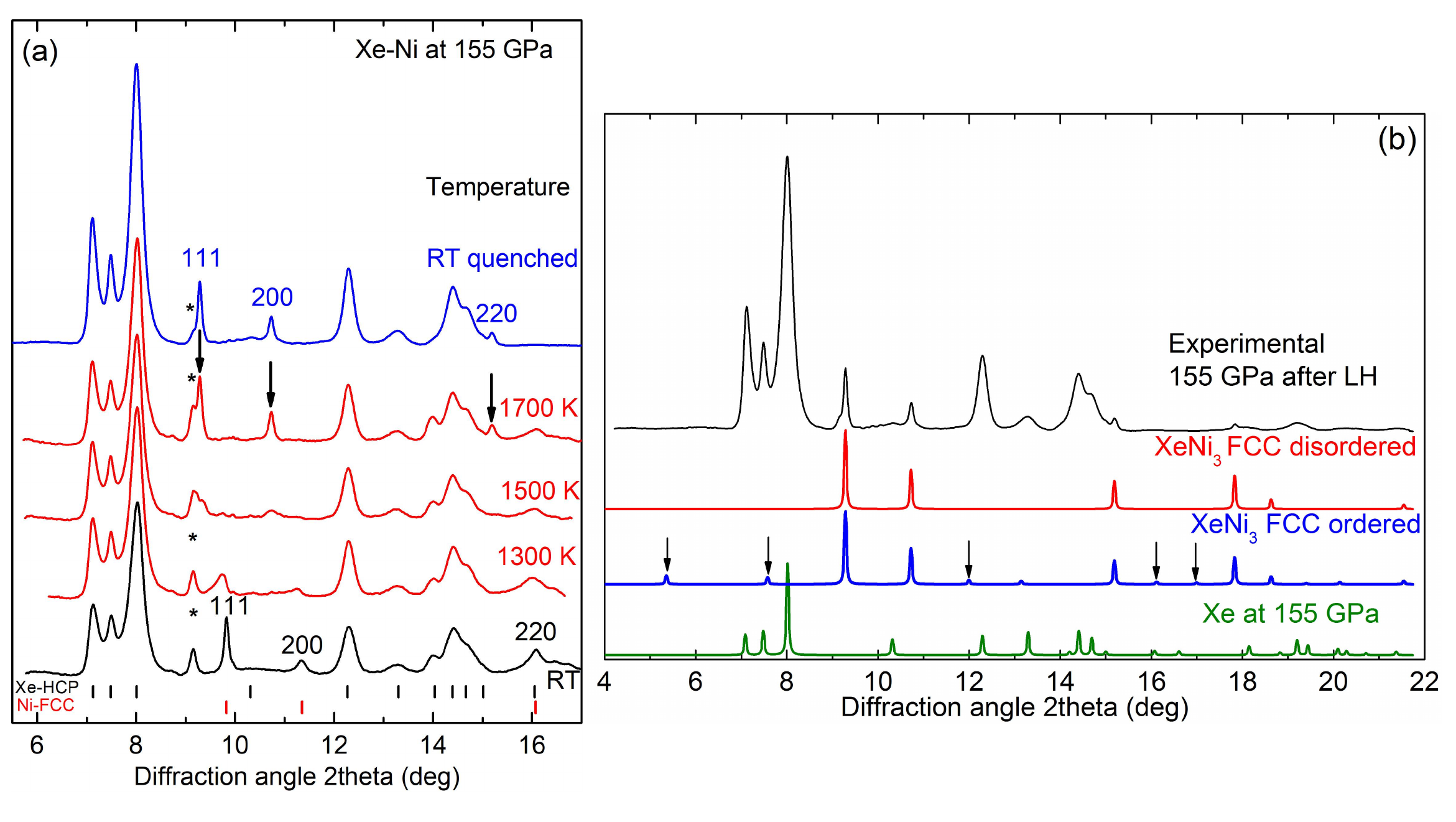}
\caption{X-ray diffraction patterns  of  XeNi$_3$. a) XRD patterns of Xe-Ni mixture at 155 GPa as a function of temperature. The peak marked by the asterisk corresponds to the strongest peak of rhenium (gasket material). The peaks of the hcp-Xe and fcc-Ni at RT before LH are marked with black and red vertical lines, respectively. The vertical arrows mark the position of the Bragg peaks of the XeNi$_3$ compound. The corresponding Miller  indices for the fcc-Ni and the fcc-XeNi$_3$   are noted at RT before and after LH respectively. b) XRD pattern of the synthesized XeNi$_3$ compound in comparison to the calculated patterns of ordered (blue) and  disordered (red) FCC crystal structures. The calculated pattern of the hcp-Xe is also shown for comparison. The vertical arrows mark the position of the additional, to the randomly substituted binary  FCC structure, Bragg peaks expected in an ordered Cu$_3$Fe-type FCC. The X-ray wavelength is 0.310 \AA.}
\end{figure}

The successful synthesis of Xe-Ni/Fe compounds in this study, well supported and corroborated  by the theoretical calculations of the present and the previous study by Zhu \emph{et al. }\cite{Zhu2014}, can be  attributed to the changing chemical properties of elements under pressure. Our results experimentally document that the metal elements gain electrons and form anions in both XeFe$_3$ and XeNi$_3$ compounds. This trend is clearly demonstrated by the calculated deformation charge density of XeFe$_3$ and XeNi$_3$, defined as the difference between the charge density of the crystal and the superimposed charge densities of non-interacting atoms (Figs. 4a and 4b). In both cases, electrons are removed from Xe (negative regions, blue) and transferred to the metals (positive regions, red).  According to previous theoretical studies (\emph{e.g.} \cite{Zhu2014, Dong2015}), application of pressure dramatically affects the chemical properties of elements. Fe and Ni in particular, become highly electronegative and can act as oxidants in compounds. Xe, on the other hand, opens up the fully filled \emph{5p} states as valence states. The charge transfer therefore takes place between the Xe \emph{5p} states and the partially filled Fe/Ni \emph{3d} or \emph{4s }(if a \emph{s} to \emph{d} transition occurs in Fe/Ni) states. Mulliken's analysis of electron density reveals the amounts of transferred charge in XeFe$_3$ and XeNi$_3$ are 0.64 e/Xe and 0.52 e/Xe, respectively, at 200 GPa. A greater amount of charge transfer in XeFe$_3$, which is visible in Fig. 4(a), is consistent with a lower occupation (\emph{d$^6$}) in the \emph{3d} states of Fe as compared to the \emph{d$^8$} occupation of Ni. The various amounts of charge transfer also likely affect the reaction pressures for these two compounds.

\begin{figure}[ht]
\centering
\includegraphics[width=130mm]{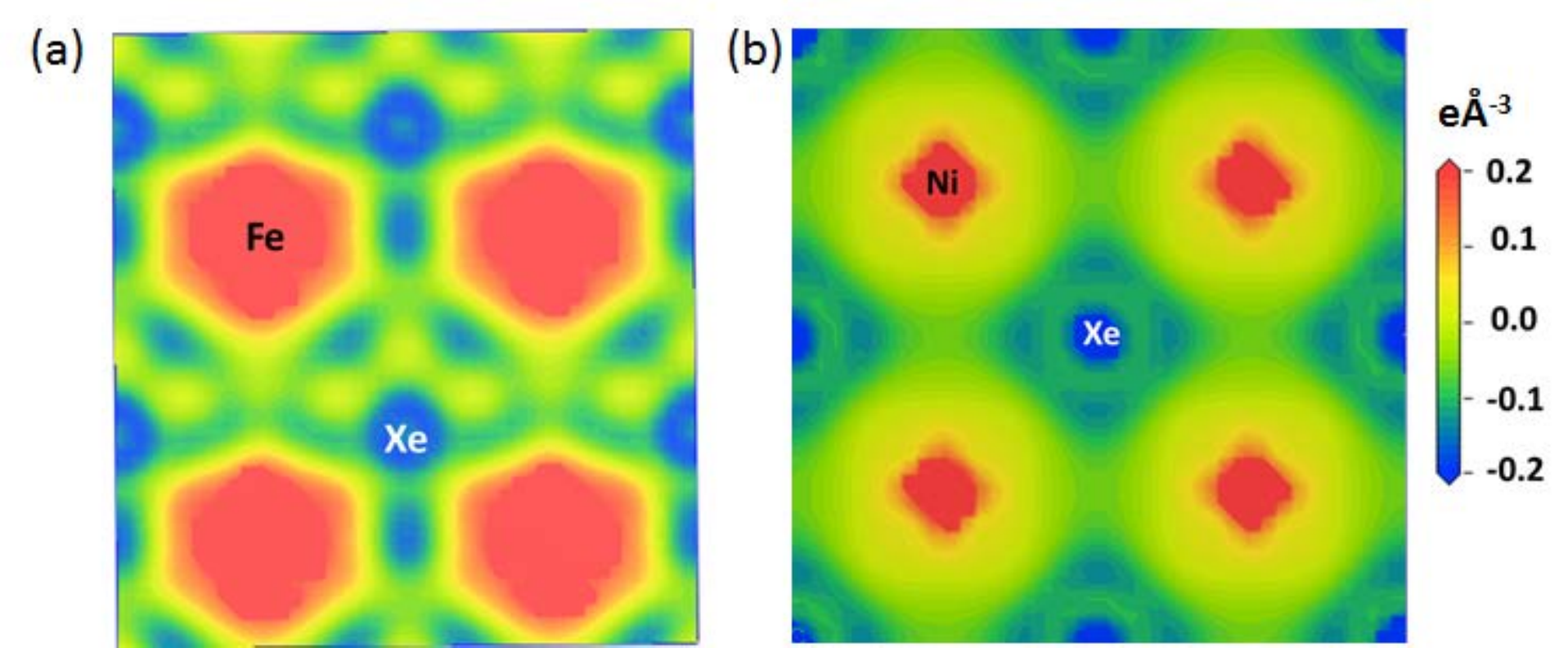}
\caption{Calculated deformation charge density of XeFe$_3$ in the (010) plane (a) and XeNi$_3$ in the (001) plane (b) at 200 GPa.}
\end{figure}

The possible formation of stable Xe-Fe compounds was previously considered as the most probable explanation of Xe depletion in the Earth's atmosphere \cite{Zhu2014,Lee2006,Nishio2010}. Although our study provides the first experimental evidence of the stability of Xe-Fe compounds at Earth's core thermodynamic conditions,  it is unlikely that such compounds have been formed during the Earth's core accretion. The formation pressure of such compounds (+200GPa), as determined in this work, is too high compared to that  suggested for the Earth's core  accretion pressure (near 50 GPa) using geochemical arguments \cite{Li1996}.   This   suggests that the formation of XeFe$_3$ is an unlikely explanation of the \textquotedblleft missing Xe paradox\textquotedblright. Thus, alternative, to the Earth's core reservoir, scenarios should be considered for the explanation of Xe depletion.

Our experiments reveal that stable compounds of metals and NGEs can exist under pressure, stabilized by a major electron transfer  from Xe to Fe (0.64e/Xe) and Ni (0.52e/Xe).   For comparison,  much lower electron transfers between Na-He (-0.174e/He) and Cs-Xe (-0.14-0.18e/Xe) were calculated in the cases of the synthesized Na$_2$He \cite{Dong2017} and predicted CsXe$_2$ \cite{Zhang2015} compounds, respectively.  This highlights a  bonding scheme that is  quite different of the ones in the cases of: a) Van der Waals  Xe-H$_2$ and Xe-N$_2$ compounds  stabilized at elevated pressures \cite{Somayazulu2010,Howie2016} and b) compounds between alkali and alkali earth metals and NGEs. This bonding pattern  resembles more the bonding between high-Z NGEs, such as Xe and Kr, and   strong electronegative elements such as F \cite{Claassen1962, Agron1963}, Cl and O \cite{Smith1963}  observed at ambient pressure. Thus, our study signifies a near halogen-like behavior of Fe and Ni  under high-pressure conditions in agreement with recent theoretical predictions \cite{Dong2015,Zhu2014}.

\section*{Conclusions}
We have performed a concomitant experimental and theoretical  search for stable compounds in the Xe-Fe/Ni system at thermodynamic conditions representative of Earth's core.  We demonstrate the synthesis of stable Xe(Fe,Fe/Ni)$_3$ and XeNi$_3$  compounds stabilized by a major electron transfer (bonding)  Xe to Fe  and Ni. Our results experimentally document that electropositive at ambient pressure elements  could gain electrons and form anions at elevated pressures.  Based on the more than 200 GPa threshold for the synthesis of XeFe$_3$ we suggest that the formation of XeFe$_3$ is an unlikely explanation of the \textquotedblleft missing Xe paradox\textquotedblright. Thus, different scenarios should be considered for the explanation of Xe depletion. Alternatively, a two step mechanism, in contrast to the direct formation of stable Xe-Fe compounds at Earth's core conditions, should be considered. In this context we may  speculate an increased solubility of Xe in molten Fe at lower accretion pressures as a first step followed by a reaction at higher pressures.  However, this extends beyond the scope of this work and calls for follow-up relevant studies.

\section*{Methods}

\subsection*{Experiments:}
High purity ($>$99.99\%) commercially available (Sigma-Aldrich) fine powders of Fe and Ni have been used for the x-ray  diffraction (XRD) measurements.  Rhenium gaskets (preindented to 30-35 $\mu$m thick using  75-50 $\mu$m beveled culets)  were used to radially confine the pressurized samples. Initial sample chamber diameters were nominally 25-50 $\mu$m. The diamond anvil cell cavity was filed with Xenon in a gas loading apparatus, where a Xe pressure of about 600psi was created, the DAC was sealed and the pressure increased to the target pressure as determined by first order Raman mode of diamond and Fe/Ni  EOSs; all readings agreed with each other within 5 GPa at pressures $>$ 200 GPa.  Integration of powder diffraction patterns to yield scattering intensity versus 2$\theta$ diagrams and initial analysis were performed using the DIOPTAS  \cite{Prescher2015} program. Calculated XRD patterns were produced  using the POWDER CELL program \cite{Kraus1996},  for the corresponding  crystal structures according to the EOSs determined  experimentally and theoretically in this study and assuming  continuous Debye rings of uniform intensity. Le Bail refinements were performed using the GSAS \cite{Larson2000} software. Indexing of XRD patterns has been performed using the DICVOL program \cite{Boutlif2004} as implemented in the FullProf Suite.

Double-sided CW laser heating was performed using ytterbium fiber lasers focused to a flat top around 10$\mu$m in diameter (FWHM) spot \cite{Prakapenka2008}. MAR-CCD  detectors were used to collect pressure dependent X-ray diffraction data at  the undulator XRD beamline at GeoSoilEnviroCARS (sector13), APS, Chicago and at Advanced Light Source, Lawrence Berkeley National Laboratory Beamline 12.2.2. The X-ray probing beam spot size was about 2-4$\mu$m at GeoSoilEnviroCARS using Kirkpatrick-Baez mirrors  and XRD data were collected \emph{in situ} at high temperature  and on the quenched samples. Temperature was measured spectroradiometrically  simultaneously with XRD measurements  with a typical uncertainty of 150 K \cite{Prakapenka2008}.   The X-ray beam was focused to 10 x 10 $\mu$m using Kirkpatrick-Baez mirrors at beamline 12.2.2. More details on the experimental set up are given in Kunz \emph{et al.} \cite{Kunz2005}. Raman studies were performed using  532  nm line of  solid-state laser and 514.5 nm line of Ar Laser in the backscattering geometry. The laser probing spot dimension was about 5 $\mu$m. Raman spectra were analyzed with a spectral resolution of 2 $cm^{-1}$ using a single-stage grating spectrograph equipped with a CCD array detector. Ultra-low frequency solid-state notch filters allowed to measure the Raman spectra down to 10 cm$^{-1}$ \cite{Stavrou2013}.

\subsection*{Theory:} Structural optimizations and total energy calculations were carried out using the Vienna \emph{ab initio} Simulation Package (VASP) \cite{Kresse1993} combined with very tight projector augmented plane-wave (PAW) potentials \cite{Kresse1999} with the Perdew-Burke-Ernzerhof (PBE) exchange-correlation functional \cite{Perdew1996}. The Fe, Ni, and Xe potentials adopt 3s$^2$3p$^6$3d$^6$4s$^2$, 3s$^2$3p$^6$3d$^8$4s$^2$, and 4d$^{10}$5s$^2$5p$^6$ as valence states, respectively. The plane-wave basis set has been carefully checked and found sufficient with an energy cutoff of 1400 eV. A 10$\times$10$\times$10, 8$\times$8$\times$6, and 5$\times$2$\times$6 k-point mesh was used, respectively, to sample the first Brillouin zone (BZ) for the FCC, \emph{Pmmn} (1) and \emph{Pmmn} (2) structures of XeNi$_3$ and XeFe$_3$. For elemental solids (Xe, Fe, and Ni), a HCP or FCC structure was used in the corresponding stable pressure regions. A 15$\times$15$\times$8 and 14$\times$14$\times$14 k-point mesh was used for the HCP and FCC structures, respectively. The new structures of XeFe$_3$ under high temperature-high pressure conditions were explored using the metadynamics method  \cite{Martonak2003,Martonak2006}combined with the VASP program. The previously predicted FCC and \emph{Pmmn} (1) structures \cite{Zhu2014} were used as initial structures for multiple simulations carried out at five pressures (180, 220, 260, 300, 340, 380 GPa) and three temperatures (2000, 3000, 4000K). Various simulation cells consisting of 8 to 24 XeFe$_3$ formula units were employed along with a k-spacing of 2$\pi$ $\times$ 0.08 \AA$^{-1}$ for BZ sampling. The Raman spectra of XeFe$_3$ were calculated using a python code \cite{Fonari2013} utilizing the macroscopic dielectric tensors determined from the zone-centered phonons calculated using the VASP program.

\section*{ACKNOWLEDGMENTS}
The authors thank Zurong Dai for assistance with SEM/EDX measurements. The authors are grateful to Sergey N. Tkachev for helping with the gas
loading at sector 13 GSECARS, which is supported by the National Science Foundation (NSF) (EAR 11-57758 and EAR-1128799), the DOE (DE-FG02-94ER14466), and
the Consortium for Materials Properties Research in Earth Sciences. Part of this work was performed under the auspices of the U. S. Department of Energy by Lawrence Livermore National Security, LLC under Contract DE-AC52-07NA27344. This work was supported by the DARPA (Grants No. W31P4Q1310005 and No. W31P4Q1210008), the Deep Carbon Observatory (DCO), and Natural Sciences and Engineering Research Council of Canada (NSERC). A. F. G. was partly supported by Chinese Academy of Sciences visiting professorship for senior international scientists (Grant No. 2011T2J20 and Recruitment Program of Foreign Experts). S.S.L. was partly supported by the Ministry of Education and Science of Russian Federation (No. 14.B25.31.0032). H. Liu was supported by EFree, an Energy Frontier Research Center funded by the DOE, Office of Science, Basic Energy Sciences under Award No. DE-SC-0001057. GSECARS is supported by the U.S. NSF (EAR-1128799) and DOE Geosciences (DE-FG02-94ER14466). Use of the APS was supported by the DOE-BES under Contract No. The Advanced Light Source is supported by the Director, Office of Science, Office of Basic Energy Sciences, of the U.S. Department of Energy under Contract No. DE-AC02-05CH11231, DE-AC02-06CH11357. Computing resources were provided by the University of Saskatchewan, Westgrid, and Compute Canada. We thank Cheng Ji, Dave Mao, and Rich Ferry for enabling the Raman measurements at HPSynC, APS.

\clearpage

\Huge\textbf{Supporting Information}

\medskip
\medskip
\medskip

\renewcommand{\thefigure}{S\arabic{figure}}
\setcounter{figure}{0}

\footnotesize\textbf{Procedure used for the identification of the \emph{Pmmn} (1) phase} : Although few of the new peaks of the XRD patterns of the  Xe-Fe$_{0.93}$Ni$_{0.07}$ mixture after LH at 210 GPa  can be indexed with a Cu$_3$Au-type structure, with a volume  in agreement with the one proposed by Zhu et al., this crystal structure does not account for the rest of the observed Bragg peaks.  Moreover, no Raman active vibrational modes are expected for  the Cu$_3$Au-type structure in contrast with our Raman spectroscopy measurements (Fig. S4). The additional Bragg peaks have been indexed, with  high figure of merit (FOM), to an orthorhombic unit cell with lattice parameters \emph{a}=5.166{} \AA, \emph{b}=10.827{} \AA {}and {} \emph{c}=3.909{} \AA {} at 220 GPa. Further analysis of the  reflection condition (systematic absences) resulted  a  primitive lattice  (P-type) with \emph{Pban }(50), \emph{Pbam }(55) and \emph{Pmmn} (59) as the most likely space groups. We examined all known theoretical models of Xe-Ni and Xe-Fe compounds with various stoichiometries \cite{Zhu2014}, but found none could satisfactorily fit  the XRD pattern.    We conducted extensive metadynamics simulations to identify possible structures of XeFe$_3$, in particular structures with competitive enthalpies to the FCC structure. Taking into consideration  both energetics and structure, a NbPd$_3$-type structure (\emph{Pmmn}, Z=6, PDF-42-1258) generated in the 200 GPa simulation   was deemed to be the best match. The calculated $\Delta$$H^f$ of the \emph{Pmmn} (1) structure is only slightly higher than the FCC structure, \emph{i.e.}, less than 0.05 eV/atom, which again indicates a close metastable phase (Fig. 2(a)). The $\Delta$$H^f$ of the FCC and \emph{Pmmn} (1) structures approaches zero at 201 and 211 GPa, respectively, both according well to the experimental reaction threshold. Our data show the presence of at least two mixed phases supporting the closeness in enthalpy of the \emph{Pmmn} and FCC phases.The positional parameters of the orthorhombic NbPd$_3$ \cite{Giessen1964} were used as the starting point for our calculations and then optimized for the XeFe$_3$ compound at 200 GPa.

\begin{figure}[ht]
\centering
\includegraphics[width=130mm]{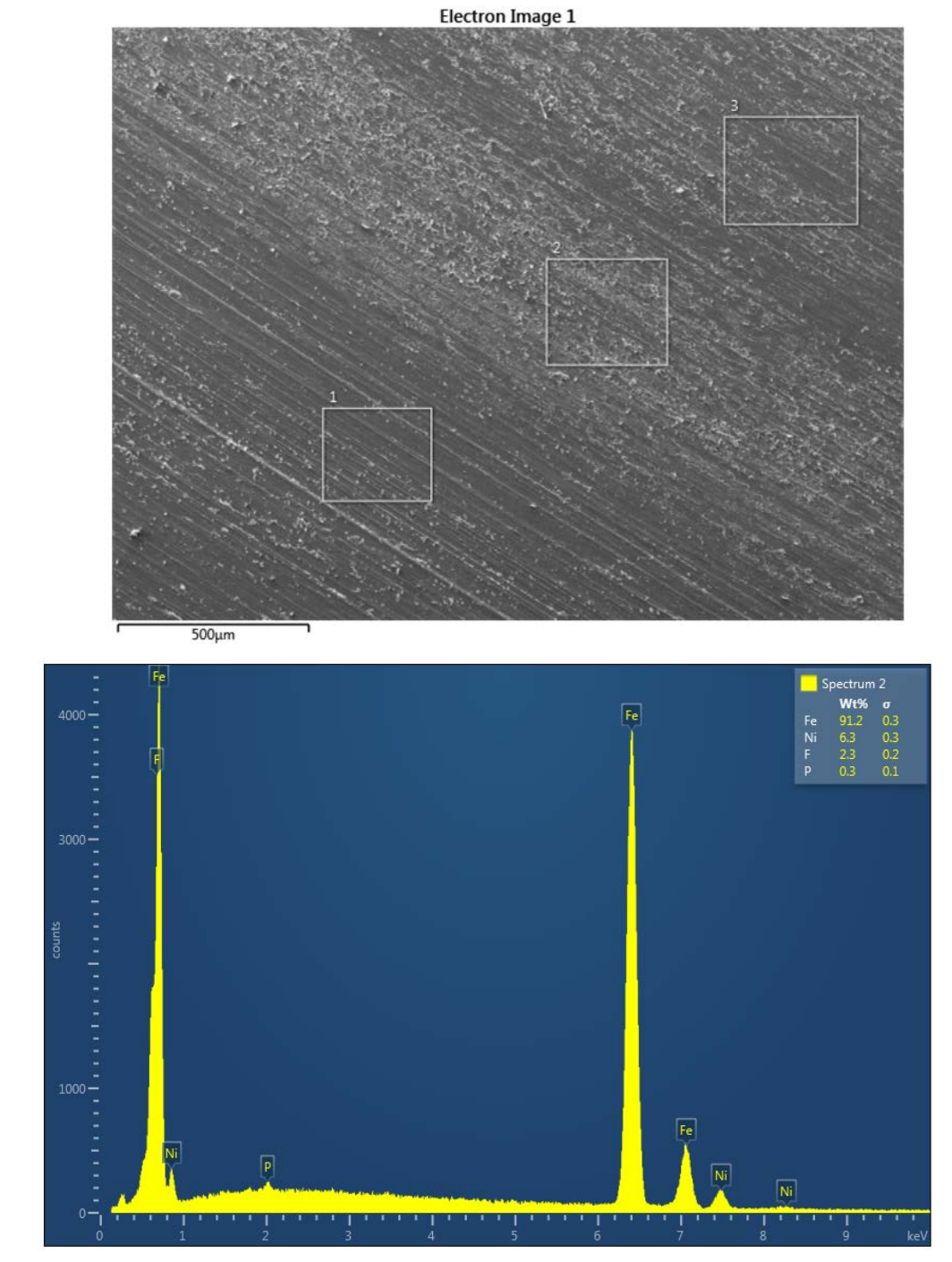}
\caption{Representative EDX spectrum of the iron Sikhote-Alin Meteorite used in this study.  Elemental composition is summarized in the Table}
\end{figure}

 \begin{figure}[ht]
 \centering
\includegraphics[width=130mm]{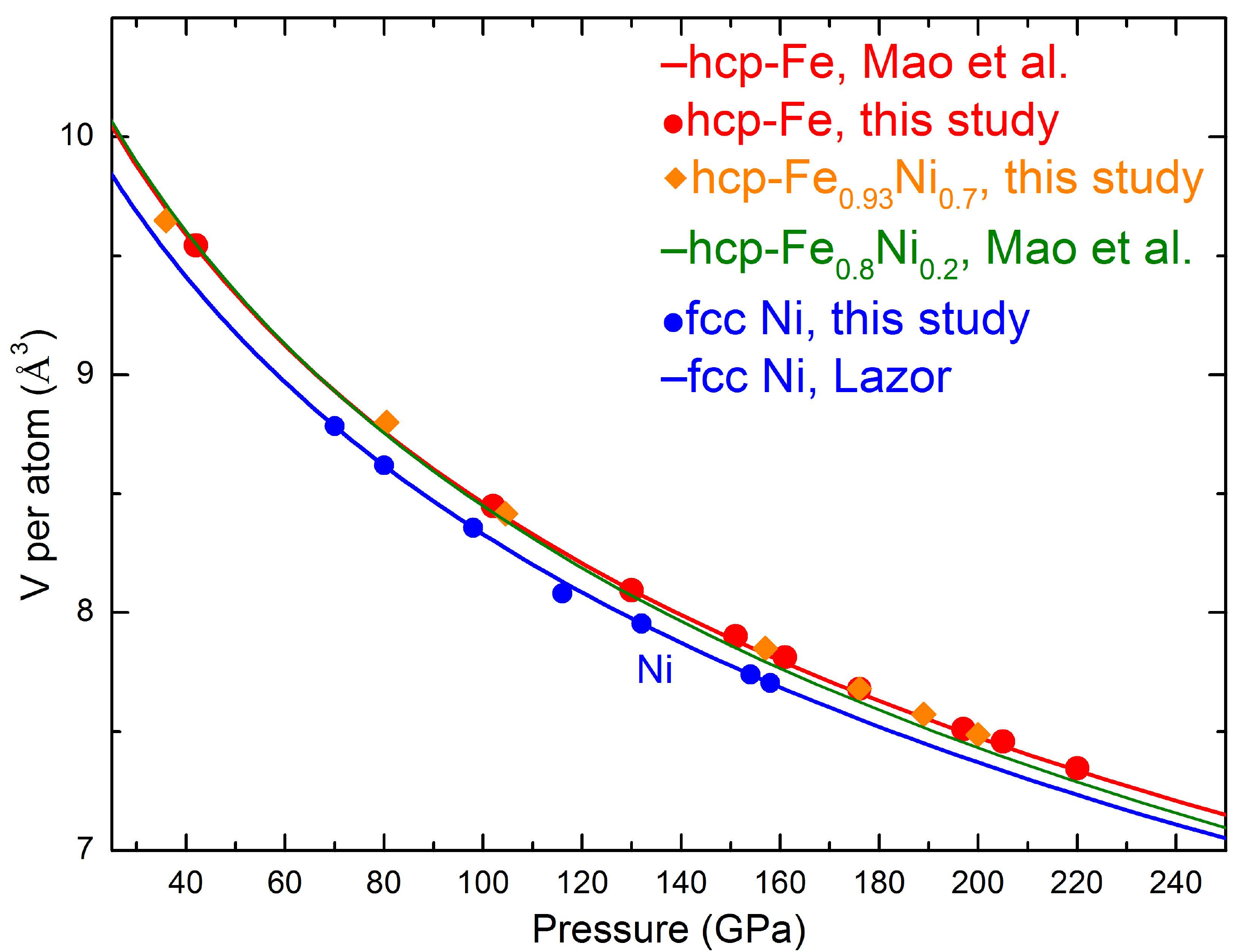}
\caption{Volume-pressure data for Fe and Fe$_{0.93}$Ni$_{0.07}$ as determined experimentally in this study. The solid lines are third-order B-M EOSs from Mao et al. \cite{Mao1990} for Fe and Fe$_{0.8}$Ni$_{0.2}$ and from Lazor \cite{Lazor1994} for Ni. }
\end{figure}

\begin{figure}[ht]
\centering
\includegraphics[width=130mm]{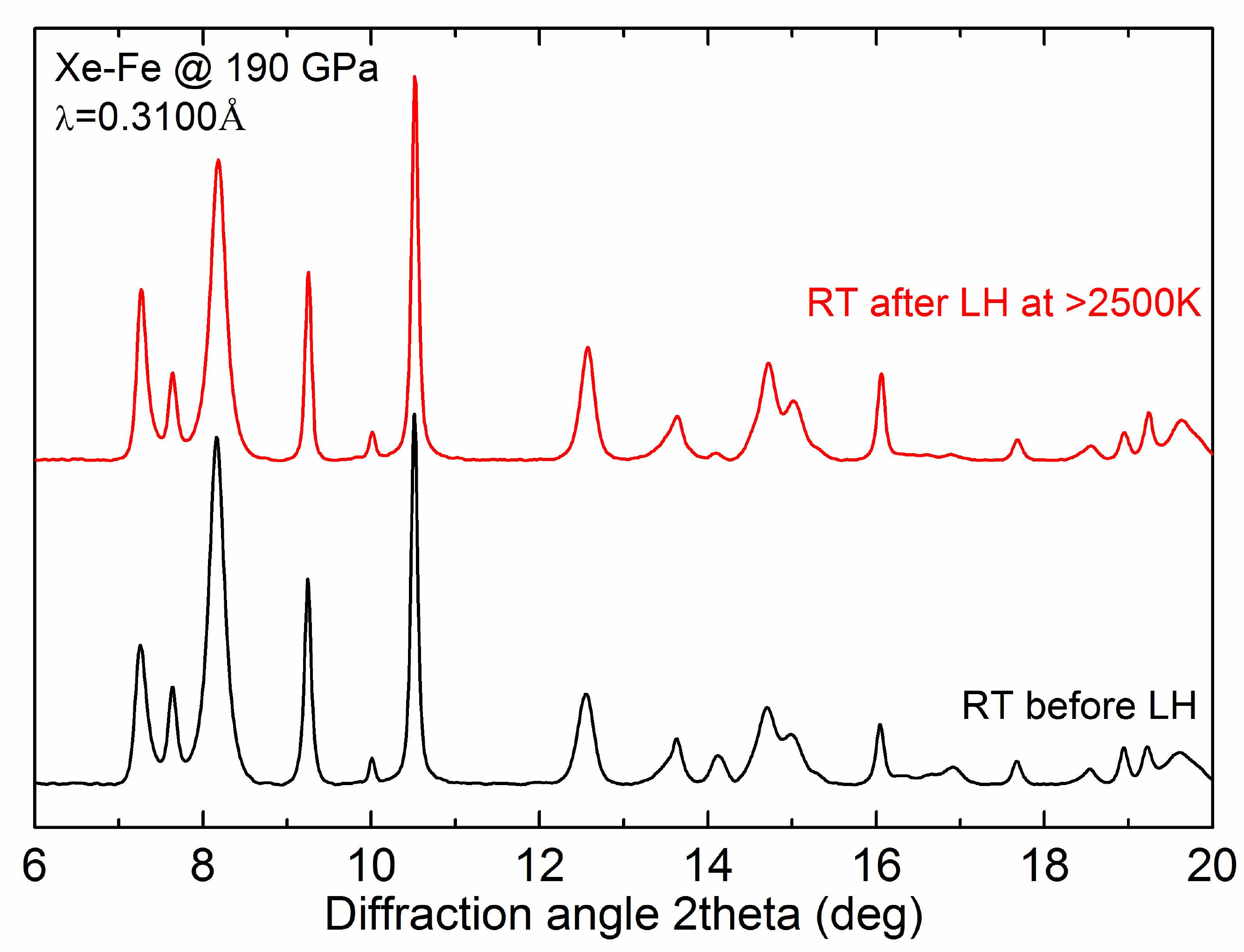}
\caption{XRD patterns of the Xe-Fe  mixture before and after LH at 190 GPa}
\end{figure}

 \begin{figure}[ht]
 \centering
\includegraphics[width=150mm]{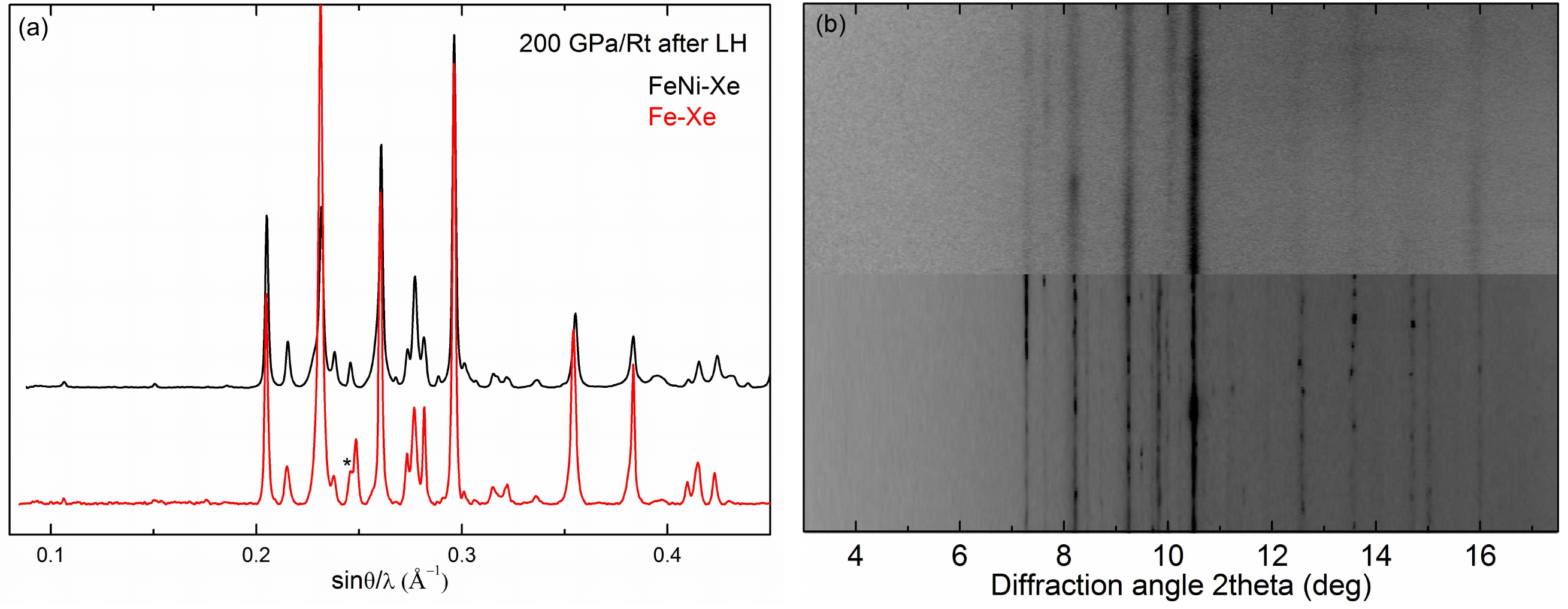}
\caption{(a) XRD patterns of the Xe-Fe and Xe-Fe$_{0.97}$Ni$_{0.07}$ mixtures after LH at 200 GPa. The peak marked by the asterisk corresponds to the strongest peak of rhenium (gasket material). The patterns are plotted as intensity versus $\sin\theta/\lambda$ (\AA$^{-1}$) \textquotedblleft normalization\textquotedblright {} because  different x-ray wavelengths (0.31\AA {} for the FeNi-Xe  and 0.3344\AA {}   for the Fe-Xe)  were used. b) 2D X-ray diffraction images in rectangular coordinates (cake) of the Xe-Fe$_{0.97}$Ni$_{0.07}$ mixture before (upper part) and after (lower part) LH at 210 GPa }
\end{figure}

 \begin{figure}[ht]
 \centering
\includegraphics[width=150mm]{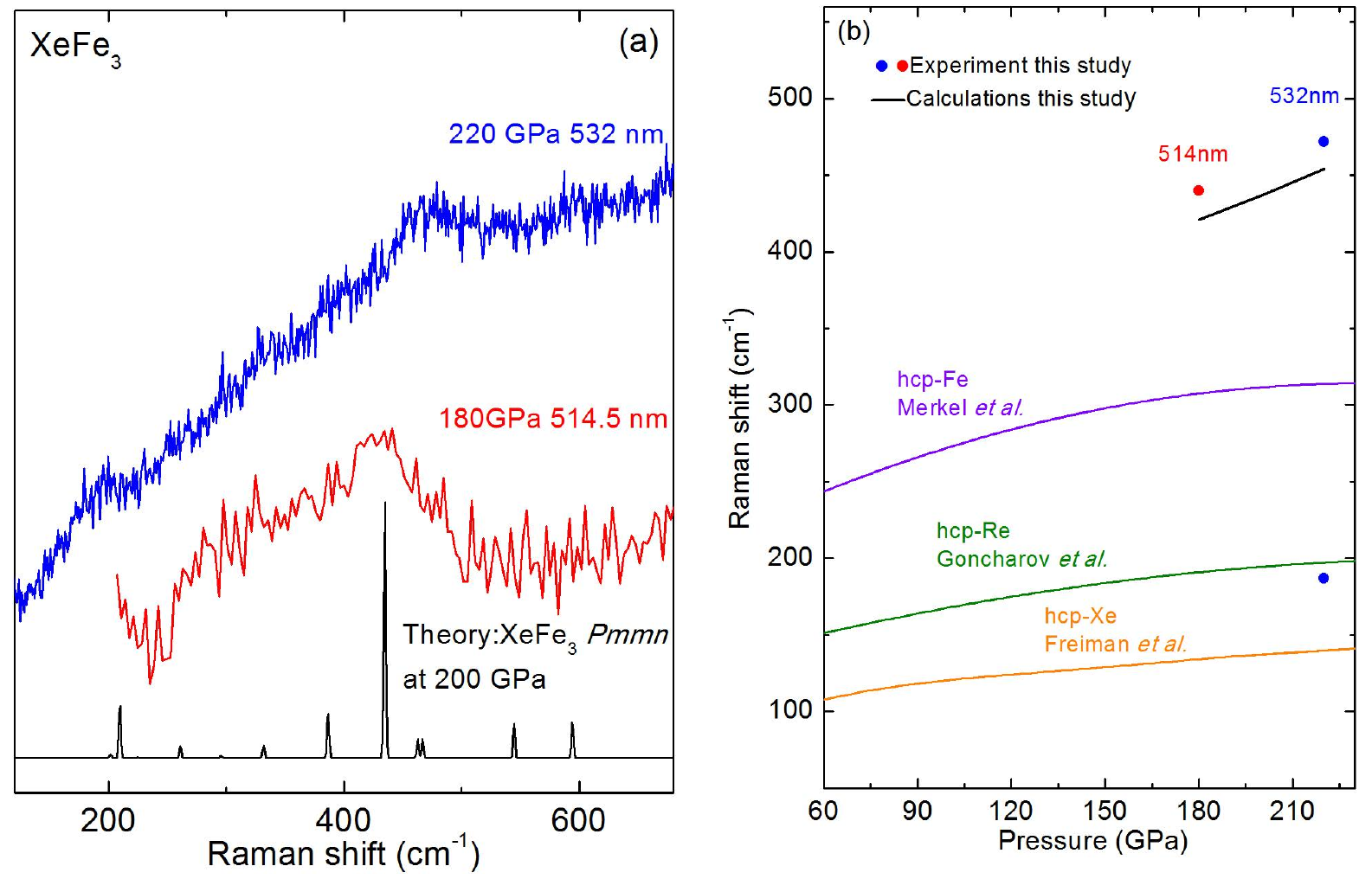}
\caption{a) Raman spectra of the Xe-Fe-XeFe$_3$ mixture after LH at 220GPa and quenched to 300 K at 220 and 180 GPa  measured with two different excitation wavelengths. Theoretically calculated Raman spectrum of \emph{Pmmn} XeFe$_3$  (blue curve) is also shown for comparison. Low intensity Raman spectra are consistent with the formation of a metallic or semi-metallic XeFe$_3$ compound \cite{Zhu2014} and consequently only  the highest intensity peak at 435cm$^{-1}$ is observed. b) Raman frequencies of the experimentally observed Raman modes (solid symbols) and of the most intense  peak of the theoretically calculated Raman spectrum (solid line)  as a function of pressure . The frequencies of the hcp-Fe \cite{Merkel2000}, hcp-Re \cite{Goncharov2002} and hcp-Xe \cite{Freiman2008} Raman modes, extrapolated at high pressures, are also plotted for comparison.}
\end{figure}

 \begin{figure}[ht]
 \centering
\includegraphics[width=130mm]{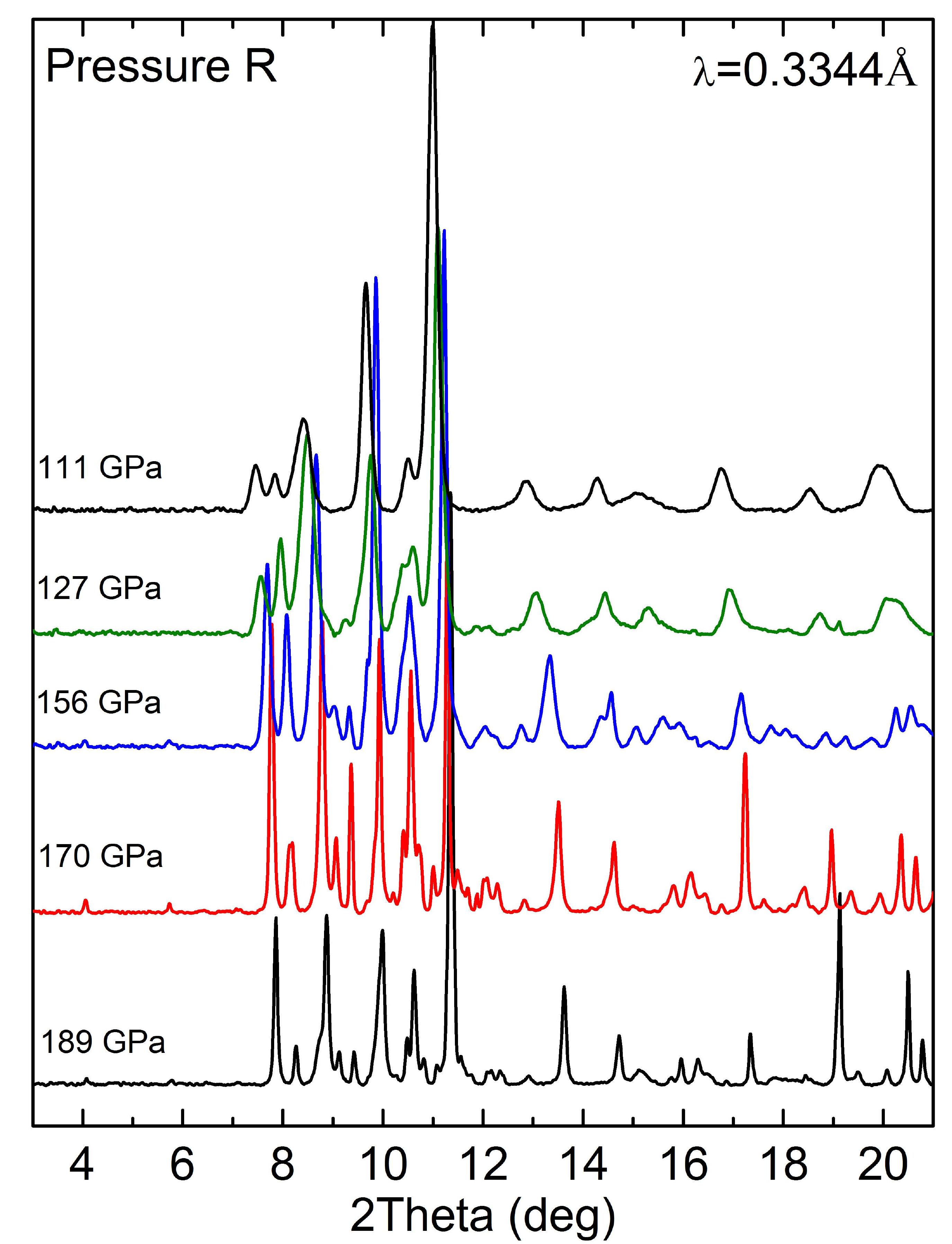}
\caption{X-ray diffraction patterns of the Xe(Fe$_{0.93}$Ni$_{0.07}$)$_ 3$, hcp-Xe and hcp-Fe$_{0.93}$Ni$_{0.07}$ mixture at various pressures on pressure release after synthesis at 210 GPa.  The X-ray wavelength is 0.3344 \AA.}
\end{figure}

\begin{figure}[ht]
\centering
\includegraphics[width=130mm]{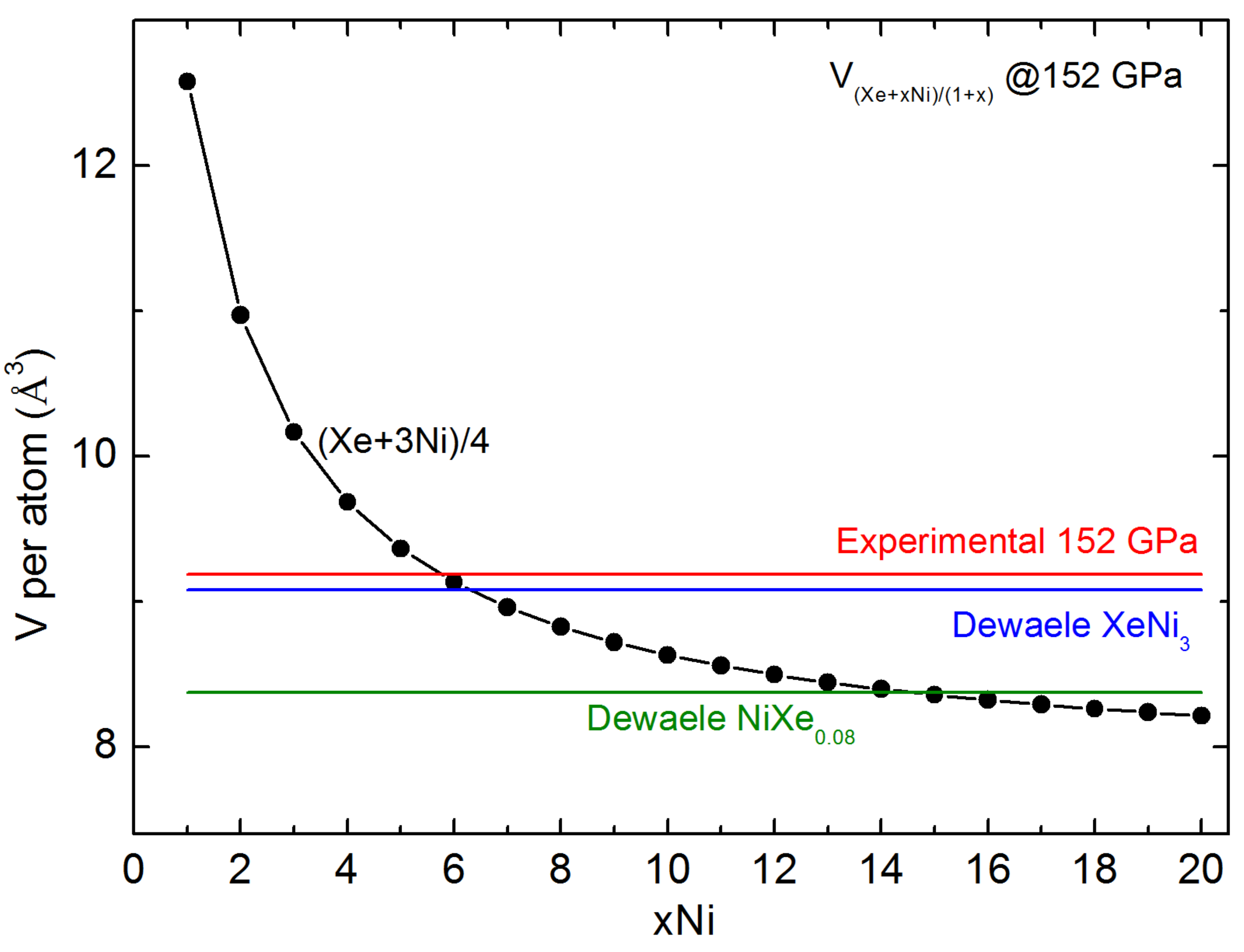}
\caption{Volume per atom of various 1:x solid mixtures of Xe and Ni, calculated according to the   experimentally determined  unit cell volumes in this study at 152 GPa. The volume per atom of the synthesized XeNi$_3$ compound at 152 GPa is noted by the horizontal red line. Results reported by Dewaele \emph{et al.} \cite{Dewaele2017} at the same pressure are also shown for comparison. }
\end{figure}

 \begin{figure}[ht]
 \centering
\includegraphics[width=155mm]{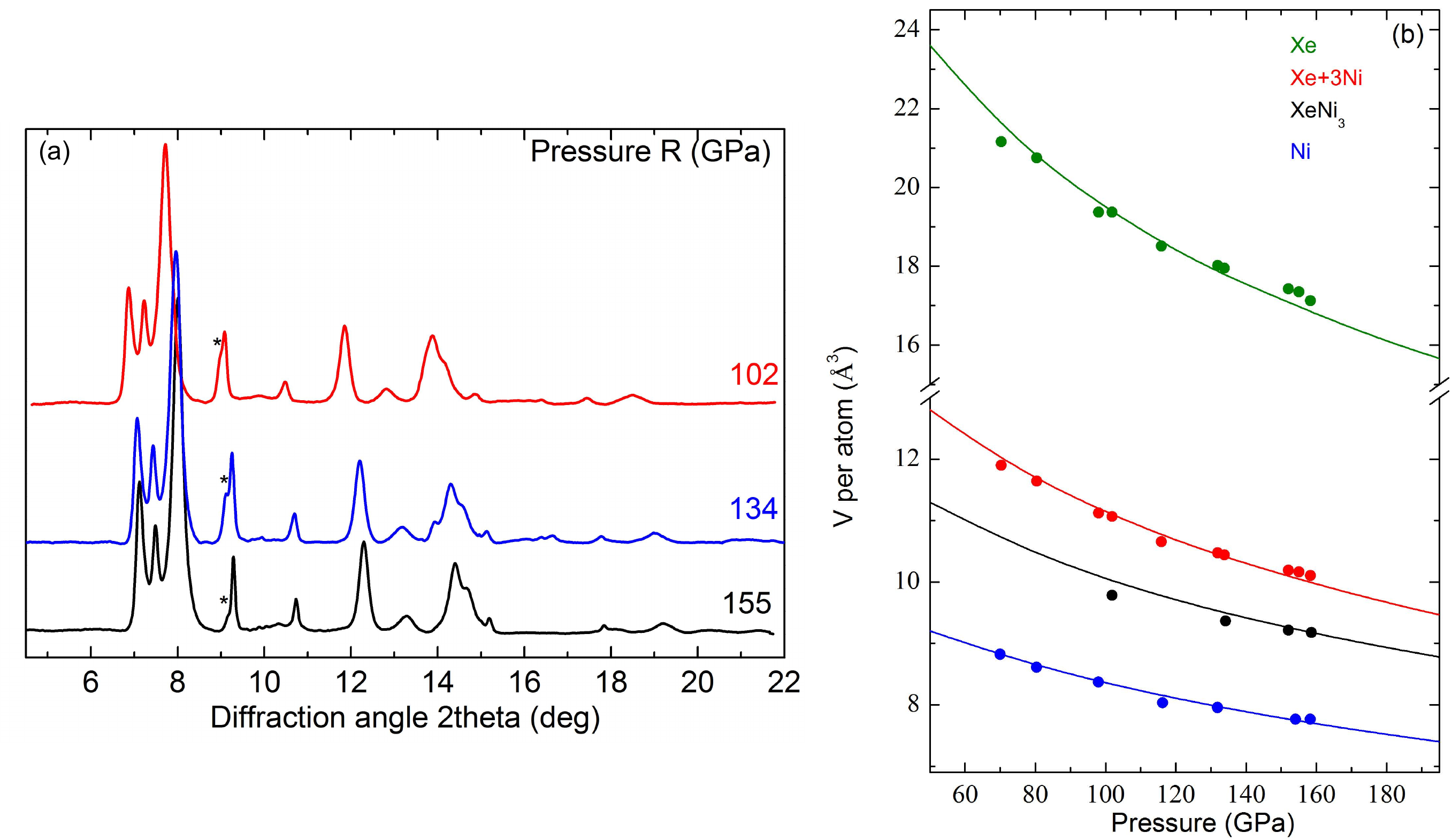}
\caption{ (a) X-ray diffraction patterns of the XeNi$_3$ and hcp-Xe  at selected pressures on pressure release after synthesis at 155 GPa. The peak marked by the asterisk corresponds to the strongest peak of rhenium(gasket material). The X-ray wavelength is 0.3100 \AA. (b)EOS of Xe, Ni and XeNi$_3$  as determined experimentally (symbols) and theoretically (lines)in this study. The volumes of XeNi$_3$ were calculated using an ordered FCC structure. The volume of the superposition of (Xe+3Ni)/4 is also shown for comparison.}
\end{figure}

\end{document}